# Derivative-free optimization of rate parameters of capsid assembly models from bulk in vitro data

Authors: Lu Xie (1 and 2), Gregory R. Smith (3), and Russell Schwartz (2 and 3)

((1) Joint Carnegie Mellon/University of Pittsburgh Ph.D. Program in Computational Biology, Pittsburgh, PA USA,

(2) Computational Biology Department, Carnegie Mellon University, Pittsburgh, PA USA,

(3) Department of Biological Sciences, Carnegie Mellon University, Pittsburgh, PA USA)

## Abstract

The assembly of virus capsids from free coat proteins proceeds by a complicated cascade of association and dissociation steps, the great majority of which cannot be directly experimentally observed. This has made capsid assembly a rich field for computational models to attempt to fill the gaps in what is experimentally observable. Nonetheless, accurate simulation predictions depend on accurate models and there are substantial obstacles to model inference for such systems. Here, we describe progress in learning parameters for capsid assembly systems, particularly kinetic rate constants of coat-coat interactions, by computationally fitting simulations to experimental data. We previously developed an approach to learn rate parameters of coat-coat interactions by minimizing the deviation between real and simulated light scattering data monitoring bulk capsid assembly *in vitro*. This is a difficult data-fitting problem, however, because of the high computational cost of simulating assembly trajectories, the stochastic noise inherent to the models, and the limited and noisy data available for fitting. Here we show that a newer classes of methods, based on derivative-free optimization (DFO), can more quickly and precisely learn physical parameters from static light scattering data. We further explore how the advantages of the approaches might be affected by alternative data sources through simulation of a model of time-resolved mass spectrometry data, an alternative technology for monitoring bulk capsid assembly that can be expected to provide much richer data. The results show that advances in both the data and the algorithms can improve model inference, with rich data leading to high-quality fits for all methods, but DFO methods showing substantial advantages over less informative data sources better representative of the current experimental practice.

# Introduction

Molecular self-assembly is by far the dominant form of chemistry in living systems, yet has been largely neglected in efforts to build predictive simulation models of complex biological systems. A substantial majority of eukaryotic proteins normally functions as parts of molecular complexes (1) and nearly every important function a cell performs – including DNA, RNA, and protein assembly and degradation, cell movement and shape depends on self-assembly of specialized structures, complexes, or molecular machines. Developing accurate, quantitative models of self-assembly processes is therefore essential to the overall mission of comprehensively simulating complex biological systems. Systems biologists have developed ever more complex and comprehensive models of biological systems (see (2) for a recent review), culminating in such recent landmarks as the simulation of whole cells (3). Yet explicit models of self-assembly reaction networks are largely absent from such efforts. This gap may reflect both the computational difficulty of handling the enormously complex networks of possible reactions produced by even simple assembly processes and the experimental difficulty of precisely monitoring the dynamics of any non-trivial molecular assembly process to properly instantiate simulations.

Viral capsid assembly has long been a key model for complex self-assembly systems. A variety of theoretical approaches have been developed to infer possible assembly pathways for such systems (4-10). Because of the large size, long time scales, and enormous space of possible pathways a large assembly might pursue (11), successful simulators require significant coarse-graining. Prevailing models accomplish this by using "local rule" models (12-15), which concisely represent a system in terms of simplified assembly subunits with sets of discrete binding sites. Locals provide a concise way to implicitly represent a potentially enormous ensemble of possible reaction trajectories by providing an efficient way to enumerate possible reactions accessible to a system from any starting state. Such local rule binding models can be combined with Brownian dynamics models (4) and/or stochastic simulation algorithm (SSA) models (16) to yield computationally tractable simulations of the assembly of potentially thousands of subunits into icosahedral capsid structures. Nonetheless, these methods have not generally been able to provide detailed quantitative models of specific capsid assemblies because they depend on detailed interaction parameters that we currently cannot measure experimentally.

Model-fitting methods provide a potential solution to this problem by allowing one to learn experimentally unobservable parameters by fitting simulations to indirect experimental measures of biological systems (17-19). Simulation-based model fitting has proven effective for a variety of simpler network models in biology (17, 20) and has previously been combined with rule-based modeling for systems that face similar problems of combinatorial blowup in pathway space to the capsid system (21-27). We have previously show that the SSA-based approach to local rule-based capsid assembly modeling is particularly amenable to assembly simulation because it greatly accelerates simulation relative to the more involved Brownian models by avoiding explicit simulations of particle diffusion, while simultaneously reducing the parameter space to a small number of kinetic parameters (16, 28, 29) that are nonetheless sufficient to capture many potential ensembles of assembly trajectories (6, 9).

Local rule capsid models are based on a notion of coarse-grained "subunits" acting as the basic building blocks for assembly. A subunit can be a single coat protein or a small tightly-bound oligomer of coat proteins, depending on system to be modeled. A local rule defines the interactions between a subunit and its neighbors via a set of discrete binding sites, each with associated geometric coordinates and kinetic interaction parameters. For an SSA variant of the local rule model, the parameters consist of on- and off-rates at the binding sites, used to instantiate a stochastic sampler over possible assembly trajectories. Using such a simulator, one can translate a simulation of assembly itself into a simulation of an experimental measure of assembly, such as static light scattering (SLS). From there, one can pose the problem of parameter inference as the inverse problem of optimizing the parameter size to minimize the deviation between true and simulated data. In prior work we demonstrated the feasibility of this approach by developing a heuristic local search algorithm (Kumar method) (30, 31) to minimize the differences between experimental SLS data and simulation outputs. The algorithm was able to learn parameters from multiple *in vitro* capsid assembly systems within experimentally plausible ranges and consistent with multiple assembly pathways types (31). This approach provided for the first time, a detailed model of subunit-level assembly for specific real viruses, giving novel insights into the unexpected complexity of interactions that may underlie true macromolecular assembly processes. In addition, it has a provided a platform for exploring how assembly may be influenced by various factors that distinguish *in vitro* models from assembly in the cell, such as macromolecular crowding (32).

While this proof-of-concept work made an important step forward for capsid assembly modeling, it also demonstrated that there are substantial obstacles to reliably inferring high-quality data fits. One important obstacle is the computational cost of assembly simulations, which can take from minutes to days for a single trajectory, depending on the parameter values because of an extensive amount of trial-and-error involved in nucleation-limited growth processes characteristic of virus assembly. A single data fitting experimental can require sampling tens of thousands of these trajectories. An even bigger obstacle is high stochastic noise, a feature inherent to the SSA modeling method. Traditional numerical optimization, such as is done in optimizing quality of fit of a parameter set, is accomplished by methods such as gradient descent, Newton-Raphson, or Levenberg-Marquardt that depend on taking derivatives of the deviation between real and simulated data. Stochastic noise in the simulated data results in discontinuities in derivatives, a significant problem for these methods. In our earlier method, the local minimization is accomplished by interpolation between fitting by gradient descent and a quadratic response surface (30, 31). While the response surface is robust against noise, gradients are highly sensitive to it. One can suppress stochastic noise by averaging a large number of trajectories, but this solution is problematic when individual trajectories are computationally costly.

To better address these challenges we here introduce the use of an alternative class of algorithms known as "derivative-free optimization" (DFO) (33). As the name suggests the class of DFO methods avoid computation of derivatives of objective function, making them in principle less susceptible to stochastic noise than are gradient methods. DFO methods in general tend to be well suited to systems such as stochastic capsid assembly, characterized by high noisy and high computational cost for evaluating the objective function. To explore the potential of DFO methods for this problem, we apply representatives of two classes of DFO method: multilevel coordinate search (MCS) (34) and stable noisy optimization by branch and fit (SNOBFIT) (35). MCS was selected because it has been reported to have the best overall performance among freely available DFO solvers (33), while SNOBFIT was chosen because it was specifically designed for optimizing noisy systems. We have applied both, along with our prior gradient-based method, to fitting parameters to real and simulated hepatitis B virus (HBV) static light scattering data. The results show that both DFO methods are able to find lower RMSDs and better parameter fits the prior methods with SNOBFIT showing the best fits.

We further sought to explore how much these questions depend on the data available for fitting. All such fitting to date has been done with static light scattering (SLS) curves (36), which measure assembly progress based on turbidity of the coat protein solution. This SLS turbidity data is experimentally simple to measure, but provides a very limited picture of assembly progress, essentially reducing the ensemble of species present at any instant of time to a single measure of average assembly size. In practice, we have learned models from SLS data by fitting simultaneously to curves derived from assembly at multiple protein concentrations to better restrict the possible parameter space. More sophisticated experimental techniques, such as time-resolved non-covalent mass spectrometry (NCMS) (37), dynamic light scattering (DLS) (38) and small angle X-ray scattering (SAXS) (39), can provide much richer measures of the assembly process with many channels of information at each instant of time. In seeking to develop the best data fitting strategies for capsids or other complex assemblies, we must consider not just improved algorithms but also how they would interact with possible improvements in experimental assays. To this end, we conducted a study of ability to learn known parameters from simulated data under three experimental conditions: a data-poor limit of fitting from a single SLS curve; a model of our current practice of simultaneous fitting to multiple SLS curves; and an idealized variant of NCMS data intended to model a best-case scenario for available data on bulk assembly in vitro. These experiments show that improvements in data do indeed lead to better model fits, but also that these improvements interact in a complex way with algorithm selection. In the data-limited case, quality of fit is highly algorithm dependent, while all three methods yield comparably high-quality fits in the limit of idealized NCMS data. The results suggest that at present, both better data and better algorithms are needed if we are to develop simulation-based data fitting as a reliable general technology for learning predictive quantitative models of complex assembly systems.

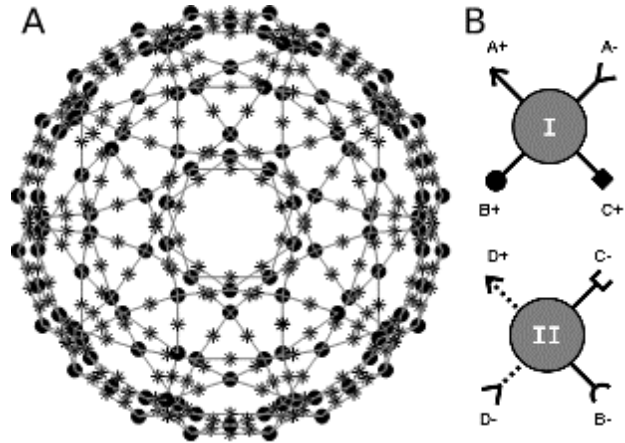

**Figure 1. HBV local rule capsid model.** (A) Model of the overall capsid geometry. Filled circles show the center positions of assembly subunits, representing coat dimers and asterisks show binding sites between adjacent subunits. (B) Local rules specifying the geometry of part (A) in terms of two dimer conformations, I and II, each with four binding sites corresponding to interactions A-D. Binding sites have specificity for with those of the same label but opposite sign: A+ to A-, B+ to B-, C+ to C-, and D+ to D-.

# Materials and Methods

## Model and Datasets

In prior work (31) we applied our custom gradient-based search algorithm (referenced below as the Kumar method) to fit models to a set of *in vitro* capsid assembly systems using SLS data, including an HBV assembly to SLS curves from Zlotnick et al (36). We use the HBV system as a model for the present work because it shows a clear nucleation-limited pathway and because it requires relatively moderate consumption of computational time per simulation trajectory. This model consists of 120 subunits, where each subunit is a dimer of coat protein. The subunits are categorized into 2 types and each type of subunit has 4 types of binding interactions describing how it can bind to neighboring subunits (Fig. 1). For each interaction type, we assign 2 kinetic parameters: an association rate (on-rate) and a dissociation rate (off-rate). We simulate trajectories from the implied model via DESSA (16), a specialized form of stochastic simulator for local rule-based self-assembly models. The simulator uses a variant of the SSA method designed to achieve linear time and space complexity per in the numbers of species present in a simulation per association or dissociation event (28), allowing it to handle the large number of intermediate species potentially present in a self-assembly trajectory, while sampling events from a Gillespie SSA model (40). In this study we use the same dataset as a test bench for two DFO methods: MCS and SNOBFIT. We use the HBV dataset because our prior work showed it to exhibit a well-defined nucleation-limited assembly pathway while yielding run times per trajectory that are computationally feasible for available cluster compute hardware.

**Table 1. On and off-rates at each binding site used in our synthetic data generation and fitting.**

| Binding site | On-rate (M$^{-1}$ s$^{-1}$) | off-rate (s$^{-1}$) | Free energy (kcal M$^{-1}$) |
|---|---|---|---|
| A | 9.48 x 10$^5$ | 7.94 x 10$^3$ | -2.83 |
| B | 5.98 x 10$^5$ | 1.26 x 10$^4$ | -2.29 |
| C | 3.78 x 10$^5$ | 2.00 x 10$^4$ | -1.74 |
| D | 2.38 x 10$^5$ | 3.16 x 10$^4$ | -1.20 |

To test the quality of fits resulting from different possible algorithms and data sources, it is necessary to have multiple data types on a common system with a known ground-truth parameter set and simulation model. Since there is, to our knowledge, no alternative to our methods for learning these properties of a complex molecular assembly, we created a synthetic variant of the HBV model using the HBV structure but a set of artificially chosen rate constants selected to maintain a realistic nucleation-limited growth mechanism while producing rapid assembly. We created synthetic datasets at four concentrations corresponding to $c$ = 5.3, 6.4, 8.0 and 10.6 µM, if we assume our simulations each represent a cubic volume of dimensions 0.5 x 0.5 x 0.5 µm$^3$ (0.125 fL). Table 1 provides the corresponding on-rates in M$^{-1}$s$^{-1}$ and off-rates in s$^{-1}$.

We generated synthetic SLS data as the source of producing static light scattering and mass spectrometry datasets. Given a simulation trajectory, we generated synthetic SLS data by computing each second the following simulated scattering intensity (41):

$$R_c(t, p_0) = k \times c \times \left( \sum_{i=1}^{n} N_i(t, p_0) \times i^2 \right) \Big/ \left( \sum_{i=1}^{n} N_i(t, p_0) \times i \right) = k \times c \times S_c(t, p_0)$$

Here $N_i(t,p_0)$ is the number of assemblies that has $i$ subunits at time point $t$, $M_i(t,p_0)$ is the mass fraction of assemblies that have $i$ subunits at time point $t$ with parameter $p_0$, $R(t,p_0)$ is the SLS intensity at time point $t$ with given parameter set $p_0$, and $n$ is the number of subunits in a complete capsid. The complete capsid consists of $n$ subunits. When fitting experimental curves, the raw curve $S(t,p_0)$ is first scaled by concentration $c$, and then scaled by a factor $k$ to match the arbitrarily unitized scattering intensity. However, for convenience in fitting synthetic data, we set $k$ such that $kc$ = 1, 1, 1.5, and 2 for concentrations $c$ = 5.3, 6.4, 8.0 and 10.6 µM, respectively.

To simulate NCMS data, we used a highly idealized model representative of the ideal theoretically possible from NCMS, assuming ability to exactly assign peaks, deconvolute contributions of distinct charge states, and precisely quantitate mass fractions at each peak. Although real data would be far noisier and more ambiguous, we believe an idealized model better serves the goal of providing a model of a maximally data-rich system. Under these assumptions, NCMS is simulated by averaging the mass fraction of each intermediate assembly (including full capsid) at every second:

$$M_i(t, p_0) = \left(N_i(t, p_0) \times i\right) \Big/ \sum_{j=1}^{n}\left(N_j(t, p_0) \times j\right)$$

For each concentration we sampled 10,000 trajectories of 250 seconds assembly time with 600 assembly subunits per simulation. Synthetic datasets used in the present work are supplied as Supporting Dataset S1.

## Optimization algorithms

We first explored the two DFO methods, MCS and SNOBFIT, on the HBV SLS dataset. In prior work we showed that the Kumar method is capable of learning physically reasonable parameters from multiple viral assembly systems (30, 31). In applying the Kumar method to real SLS data, we combined its gradient/quadratic response surface local optimizer with a scheme for successively subdividing parameters to create an efficient heuristic global optimizer over the log parameter space. In order to achieve comparable results by the DFO methods, we adopted the same parameter grouping and splitting scheme in applying the two DFO methods to the real HBV data. The MCS solver treats evaluations of quality of fit as black-box function. At each stage of the scheme, it calls a wrapper for the DESSA simulator with different sets of parameters and expects return values in the form of the measure of fit to be optimized. For each set of parameters, the return value is evaluated by averaging 50 replica trajectories with MCS solver deciding when to terminate the search of current stage. The SNOBFIT solver works uses a different interface that allows it to be used as a direct replacement of Kumar solver. At the beginning of each stage, we send SNOBFIT fit function values at the same initial grid points as used in the initial setup of the Kumar method, and query the SNOBFIT solver for an equal number of points in log parameter space to evaluate in the next iteration. Among the new points, half of them are best-guess predictions of the fit minima, and the other half are deployed to explore the rest of the log parameter space. The function value at each point is evaluated by averaging 50 simulation trajectories. We repeat the process of iteratively feeding function values to SNOBFIT and asking it to generate for same amount of new points until there is no improvement for 10 consecutive iterations.

In addition to feeding function values, we also need to provide SNOBFIT with information about the noisiness of the function values. For simplicity we did not estimate the stochastic noise for every iteration, but rather estimate noise at one point by bootstrapping and use that noise estimate estimation as a fixed value for every point sampled. The bootstrapping estimation was been done in two ways. For experimental data, since we do not know the true parameters, we performed bootstrapping with the initial parameter set. We simulated 1,000 trajectories with the initial parameter set and randomly picked 50 trajectories to compute the RMSD. We repeated that for 100 replicates with replacement and treated the standard deviation of the RMSDs as an estimate of data noise. For synthetic datasets, since we know the true parameters, we did bootstrapping with the true parameter set. We simulated 10,000 trajectories with the true parameter set and randomly picked 50 trajectories to compute the RMSD, sampling 100 times with replacement and treating the mean of the RMSDs as an estimate of data noise. Comparison of noise estimates from 1,000 trajectories obtained from the initial parameter set and bootstrapping with synthetic SLS data revealed a 20% difference in estimates, suggesting that it is reasonable to apply a common noise value to all points evaluated along the search invoked by SNOBFIT.

The procedure was somewhat altered for evaluation on synthetic data. The fast simulation of synthetic datasets make it feasible to search directly in the 8-parameter space (on- and off-rates for each of four binding sites) with more points evaluated every iteration, rather than using the parameter subdivision scheme we applied to the real data. For the Kumar method, on each search iteration, the fit objective function is evaluated at 128 grid points in the log parameter space and a new minimum candidate is predicted based on the objective values. The grid points are picked in the same fashion as in the previous work (31). The function value is evaluated by averaging 40 replica trajectories at each grid point and 1280 replicas at the minimum candidate to minimize stochastic noise. On average, each point is evaluated by (128*40+1280)/(128+1)≈49.6 trajectories, which is close in number to the 50 trajectories used for evaluating each point in MCS and SNOBFIT. MCS is again used as a caller for our simulator, with the MCS solver determining which parameter points to evaluate and when to terminate the search. SNOBFIT is fed with values at 128 points and asked for 128 new points to evaluate each iteration, with the search terminating after 10 non-improving iterations.

We explored fits to three variants of the synthetic data: 1) 1-SLS, where we only fit the synthetic light scattering curve with $c$ = 8.0 µM; 2) 3-SLS, where we fit three synthetic light scattering curves with $c$ = 5.3, 8.0 and 10.6 µM simultaneously; and 3) MS, where we fit the synthetic mass spectrometry dataset with $c$ = 8.0 µM. The parameters inferred by fitting synthetic datasets were then used to predict the assembly behavior under concentration $c$ = 6.4 µM, so we can investigate the deviation between quality of fit at concentrations found in the training data and those projected by the simulator using parameters learned at other concentrations.

The objective function for fitting SLS data is the root mean square deviation (RMSD) between the target SLS curve and the curve we want to evaluate:

$$f_{m-SLS}(p) = \sqrt{\frac{1}{m}\sum_c \frac{1}{T}\sum_t \left(R_c^*(t,p) - R_c(t,p_0)\right)^2}$$

Here $T$ is the number of total time points and $R^*$ is the curve obtained at the point we want to evaluate, and $m$ is number of concentrations measured in the dataset. In the 1-SLS fitting case $m$=1 and in the 3-SLS fitting case $m$=3. The curves to be fit, $R_c$, can either be synthetic curves or real experimental measured curves (with unknown $p_0$).

When fitting MS data the objective function is the RMSD between all pairs of mass fraction curves:

$$f_{MS}(p) = \sqrt{\frac{1}{n}\sum_{i=1}^{n} \frac{1}{T}\sum_t \left(M_i^*(t,p) - M_i(t,p_0)\right)^2}$$

Here $M_i^*$ is the mass fraction of the assembly with size $i$ from evaluating the the point at $p_0$. Since the curves for different datasets have distinct heights, similar amount of deviation may reflect larger difference in RMSD values. To provide a fair comparison across datasets, we normalize the RMSD obtained in each optimization search by the root mean square height (RMSH) of its respective dataset:

$$F_{m-SLS}(p_0) = \sqrt{\frac{1}{m}\sum_c \frac{1}{T}\sum_t R_c(t,p_0)^2}$$

Table 2. Qualities of fit to real HBV SLS data for the Kumar, MCS, and SNOBFIT methods.

| Method | Kumar | MCS | SNOBFIT |
|---|---|---|---|
| On-rate ($M^{-1} s^{-1}$) | $1.40 \times 10^6$ | $1.24 \times 10^6$ | $8.90 \times 10^5$ |
| A off-rate ($s^{-1}$) | $1.20 \times 10^5$ | $9.86 \times 10^4$ | $5.20 \times 10^4$ |
| B off-rate ($s^{-1}$) | $1.40 \times 10^5$ | $1.15 \times 10^5$ | $1.90 \times 10^5$ |
| C off-rate ($s^{-1}$) | $1.40 \times 10^5$ | $1.15 \times 10^5$ | $2.50 \times 10^4$ |
| D off-rate ($s^{-1}$) | $1.20 \times 10^5$ | $9.91 \times 10^4$ | $9.30 \times 10^4$ |
| A Free energy (kcal $M^{-1}$) | -1.45 | -1.50 | -1.68 |
| B Free energy (kcal $M^{-1}$) | -1.36 | -1.41 | -0.91 |
| C Free energy (kcal $M^{-1}$) | -1.36 | -1.41 | -2.12 |
| D Free energy (kcal $M^{-1}$) | -1.45 | -1.50 | -1.34 |
| #Points | 886 | 1615 | 1229 |
| Minimum RMSD | 0.0591 | 0.0486 | 0.0475 |

The columns, in order, provide inferred on-rate (assumed equal for all sites); inferred off-rates for binding interactions of types A, B, C, and D; number of function evaluations required; and the RMSD of the best-fit parameters.

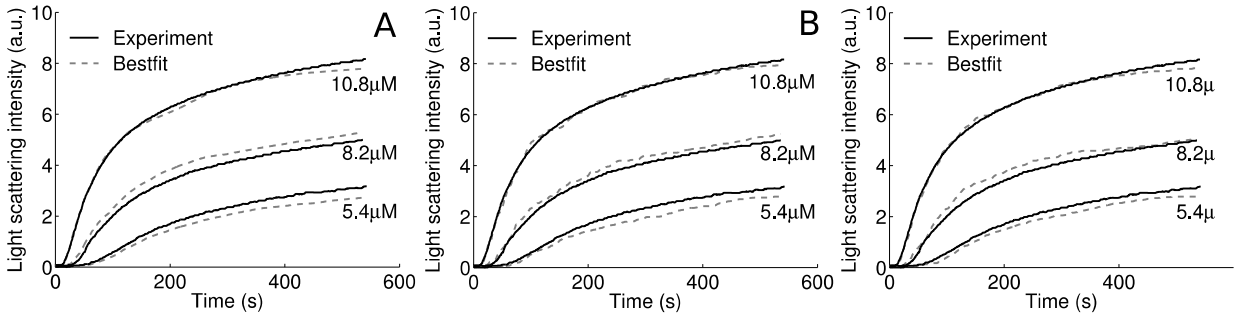

Figure 2. Best fit curves for each method to the real HBV SLS data (36) for (A) Kumar; (B) MCS; and (C) SNOBFIT. Each subpart shows three concentrations used in fitting, with true data in solid black lines and the best model fits in dashed grey lines.

$$F_{MS}(p_0) = \sqrt{\frac{1}{n}\sum_{i=1}^{n}\frac{1}{T}\sum_{t}M_i(t,p_0)^2}$$

The RMSH is equivalent to the RMSD between the curves in the dataset and a zero line.

Source code for the DESSA self-assembly simulation program, which was used for stochastic simulations in the present work, can be downloaded from the following URL: http://www.cs.cmu.edu/~russells/software/Dessa_1_5_8.zip. MCS (34) and SNOBFIT (35) are available from their authors, as described in the relevant citations.

# Results

We first examined data fits on the real HBV SLS data for each of the three methods: the gradient-based Kumar method from our prior work and the DFO methods MCS and SNOBFIT. Tests in this case were restricted to experimental data (36) and used the parameter-subdivision scheme from our prior work

Table 3. Assessment of synthetic data fitting for the Kumar, MCS, and SNOBFIT methods.

| Method | Dataset | #points | Mean parameter deviation | Best fit RMSD | Prediction RMSD (SLS) | Prediction RMSD (MS) |
|---|---|---|---|---|---|---|
| Kumar | 1-SLS | 2710 | 2.01 | 0.0667 | 0.981 | 1.01 |
| MCS | 1-SLS | 2021 | 1.48 | 0.0325 | 0.174 | 0.716 |
| SNOBFIT | 1-SLS | 2322 | 0.85 | 0.0161 | 0.0545 | 0.113 |
| Kumar | 3-SLS | 4645 | 1.99 | 0.0781 | 0.132 | 0.766 |
| MCS | 3-SLS | 2361 | 1.62 | 0.0572 | 0.123 | 0.613 |
| SNOBFIT | 3-SLS | 3225 | 0.46 | 0.0188 | 0.0166 | 0.0180 |
| Kumar | MS | 3097 | 0.34 | 0.0457 | 0.0558 | 0.0502 |
| MCS | MS | 1657 | 1.11 | 0.0296 | 0.0929 | 0.0533 |
| SNOBFIT | MS | 3741 | 0.88 | 0.0198 | 0.0133 | 0.0138 |

The columns, in order identify the data source, number of functional evaluations required, mean error in parameter fits, RMSD of the best-fit parameters to the training data, RMSD of the best-fit parameters to SLS data at a concentration not used in training, and RMSD of the best-fit parameters to NCMS data at a concentration not used in training.

Table 4. Parameter deviations between inferred and true binding rates fit to synthetic data fitting by the Kumar, MCS, and SNOBFIT methods.

| Method | Dataset | A-on | B-on | C-on | D-on | A-off | B-off | C-off | D-off |
|---|---|---|---|---|---|---|---|---|---|
| **Kumar** | **1-SLS** | -1.13 (0.10) | 0.74 (0.05) | 0.65 (0.04) | -1.70 (0.12) | 3.41 (0.12) | 3.22 (0.13) | 1.47 (0.09) | 3.74 (0.14) |
| **MCS** | **1-SLS** | 0.00 (0.52) | 2.17 (0.40) | 1.29 (0.18) | 1.75 (0.04) | 2.71 (0.34) | 1.28 (1.10) | 0.97 (0.72) | 1.68 (0.47) |
| **SNOBFIT** | **1-SLS** | 1.98 (0.96) | 0.60 (1.08) | -0.49 (0.92) | -0.28 (0.80) | 1.42 (0.86) | 0.22 (0.86) | 0.30 (0.79) | 1.51 (1.14) |
| **Kumar** | **3-SLS** | -0.75 (0.10) | 2.01 (0.09) | 2.61 (0.13) | -0.25 (0.08) | 0.97 (0.07) | 1.65 (0.08) | 2.83 (0.12) | 4.87 (0.22) |
| **MCS** | **3-SLS** | 0.13 (0.24) | 2.90 (0.69) | 0.70 (0.37) | 1.83 (0.14) | 2.29 (0.23) | 2.83 (0.88) | -0.70 (1.23) | 1.59 (0.95) |
| **SNOBFIT** | **3-SLS** | 0.13 (1.21) | -0.01 (1.12) | 1.54 (0.99) | -0.75 (0.59) | 0.15 (0.94) | -0.21 (0.60) | 0.61 (0.64) | -0.16 (0.90) |
| **Kumar** | **MS** | -0.14 (0.02) | -0.21 (0.02) | -0.35 (0.02) | 0.83 (0.06) | -0.45 (0.04) | -0.28 (0.04) | -0.10 (0.04) | 0.36 (0.03) |
| **MCS** | **MS** | -0.84 (0.02) | 2.84 (0.03) | 0.64 (0.11) | -0.32 (0.12) | -1.10 (0.10) | 3.10 (1.63) | 0.01 (0.49) | 0.01 (0.17) |
| **SNOBFIT** | **MS** | 1.06 (0.57) | -0.98 (0.76) | 1.96 (1.43) | -0.73 (0.61) | 0.26 (0.60) | 0.06 (0.25) | 1.47 (1.19) | -0.50 (0.52) |

Each entry provides the log deviation between the best-fit and ground-truth parameter; the values in parentheses are standard deviations of the fit, estimated from sub-optimal parameters that yield objective values within 2 standard deviations of the best-fit.

(30, 31) which leads to inference of a single on-rate and four distinct off-rates. Table 2 summarizes the inferred parameter values and quality of fit for each method, as assessed by RMSH. Fig. 2 plots the best-fit SLS curves for true and inferred models by each method. The figure shows superficially similar qualities of fit for each method, with good overall fits to each curve but in general a slight overprediction of the middle concentration and underprediction of the upper and lower concentrations. The table shows similar parameter inferences as well for the methods, with the most extreme outlier being a variation of 5.7-fold between inferred C off-rates between the Kumar and SNOBFIT methods. The RMSD suggest that the quality of fits are similar, with SNOBFIT performing best, followed by MCS, then Kumar. The methods also show variations in numbers of points they required to make a fit, with Kumar most efficient by this measured, followed by SNOBFIT, then MCS.

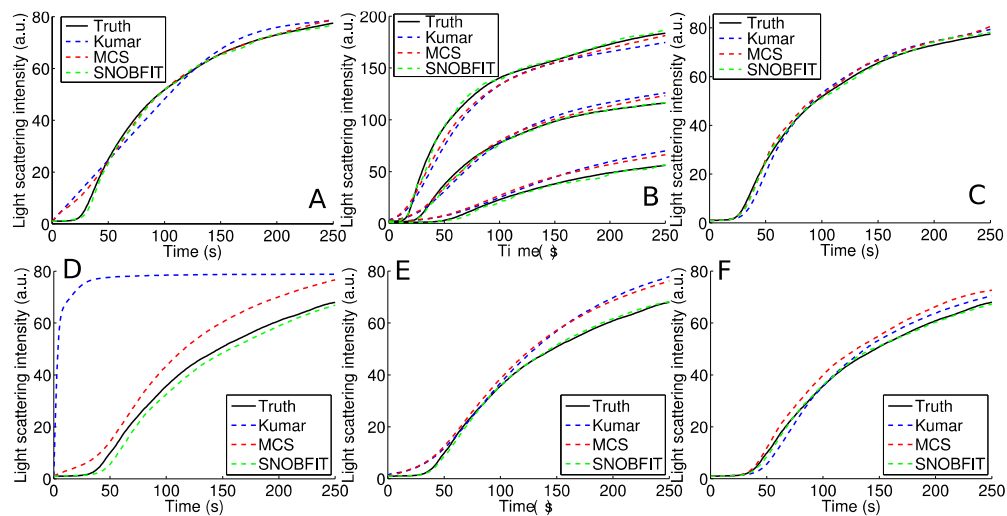

**Figure 3. Best fit curves for each method to the synthetic data as assessed by light scattering intensity.** (A) 1-SLS at the concentration to which the model was fit; (B) 3-SLS at the concentrations to which the model was fit; and (C) NCMS at the concentration to which the model was fit; D) 1-SLS at a concentration not used in fitting; (E) 3-SLS at a concentration not used in fitting; and (F) NCMS at the concentration at a concentration not used in fitting. Each subpart shows the curve simulated from the true parameters in solid black lines and the best fits for the Kumar method in dashed blue, MCS in dashed red, and SNOBFIT in dashed green.

We next examined the performance of the various methods on simulated data with a known ground-truth parameter set. Comparing the deviation of inferred parameters (Table 3) across different datasets and methods, we can conclude that SNOBFIT gives the most consistent estimates across data types, while Kumar method is most dependent on the quality of the data source. SNOBFIT consistently gives the lowest best-fit RMSD, while the Kumar method gives the highest RMSD. The difference in best-fit RMSD across the three datasets is small, however. All three methods make much better predictions, in terms of RMSD, from fitting to the richer datasets.

Given that small changes in RMSD can correspond to significant changes in inferred parameters, we extended our consideration of fit quality to consider not just the best-fit but also the range of fits within the margin of noise of the best-fit for each method and data source. In Table 4, we provide deviations between best-fit and true parameters for each method, data source, and rate parameter as well as standard deviations of these fits across the set of near-optimal parameter sets, defined as those with RMSD score within two standard deviations of the best-fit. Supplementary Fig. S1 provides box plots showing in more detail the ranges of inferred parameter values across these near-optimal parameters. The standard deviation here is the same quantity we used to estimate noise levels for SNOBFIT by bootstrapping. The Kumar method gives the tightest distribution of near-optimal parameter sets, which results from the nature of its local optimization strategy. SNOBFIT gives the widest distribution of near-optimal parameter sets. Fitting richer datasets tightens the distribution of near-optimal parameter sets for each method, showing that the greater complexity of data is helpful in more precisely pinning down the true optimal fits.

We next examined the quality of fits in terms of true and inferred profiles of assembly progress versus time, which provide a more direct view of fit quality. Fig. 3 compares fits to simulated SLS profiles of each data set used in fitting. For the 1-SLS data (Fig. 3A), only SNOBFIT gives a close fit to the experimental curve, with best-fits from Kumar and MCS showing substantially shorter lag phases and less pronounced sigmoidal behavior than the true curve. 3-SLS (Fig. 3B) leads to an improvement for all three methods, although SNOBFIT still yields noticeably better fits than the others. All three methods give good fits to the NCMS data (Fig. 3C), with apparently similar fit quality for the three.

A more stringent test is given by examining fits to mass fractions of specific intermediate species, a test examined in the Supplementary Material in Fig. S2. For this purpose, we picked three sizes of assemblies as representatives of three ranges of abundance, as assessed by mass fraction: full capsid (high abundance), trimer of subunits (medium abundance), and decamer of subunits (low abundance). As expected, the algorithms all do a better job fitting more abundant species, which is unsurprising since those species would have greater weight in computing the objective function. The high abundance capsomer species (Fig. S2, right column) is fit well by SNOBFIT for all data sources but by Kumar and MCS only when fit to NCMS data. The medium abundance trimer intermediate (Fig. S2, left column) is fit poorly by all three methods on 1-SLS data, very well only by SNOBFIT for 3-SLS data, and fit well by all three methods for NCMS data. The low-abundance decamer species i (Fig. S2, middle column) is poorly fit by all three methods, with the inferred curves deviating from the true curves by roughly a factor of two for all three methods on NCMS data.

An even more stringent test, to help control for the possibility of overfitting, is to evaluate fit quality at an additional concentration not used in parameter inference. In relative concentrations, this experiment involves learning fits from $c$ = 8.0 µM for 1-SLS and NCMS data or from $c$ = 5.3, 8.0, 10.6 µM for 3-SLS data, then evaluating the quality of the fit in each case at $c$ = 6.4 µM. Fig. 3D-F shows qualities of fit for simulated SLS curves. For 1-SLS (Fig. 3D), only SNOBFIT yields a high quality prediction, with moderate but noticeably worse quality for MCS and very poor fitting for the Kumar method. Prediction from 3-SLS data (Fig. 3E) similarly shows a high-quality fit only for SNOBIT. Kumar and MCS show nearly identical fits to one another, with the Kumar fit substantially better than it was for 1-SLS but the MCS fit very similar to that found by MCS on 1-SLS data. All methods give plausible fits with parameters inferred from fitting NCMS data (Fig. 3F), although the SNOBFIT curve is still slightly better than those from Kumar and MCS.

The pattern seen in predicted mass fractions for a concentration not used in fitting (Fig. S3) is similar to that found in Fig. S2 testing fit to the training data. The high-abundance capsomer species is fit well only by SNOBFIT for 1-SLS and 3-SLS data, but by all three methods for NCMS data. Fitting is poorer for the medium-abundance trimer species for all methods, but still reasonable for all three with NCMS data, good only for SNOBFIT with 3-SLS data, and poor for all three methods with 1-SLS data. None of the methods achieves a close fit to the low-abundance decamer species from any data set, although all come within approximately a factor of two for NCMS data.

# Discussion

This study has examined the potential of simulation-based data-fitting applied to bulk in vitro kinetic data as a technology for learning detailed models of complex self-assembly reactions, using virus capsid assembly as a model. Capsid assembly makes an excellent model for such questions because it as an exceptionally difficult system for such methods due to its large size and space of possible pathways but has also intensively studied by experiment, theory, and simulation. Our results suggest that learning accurate model parameters and assembly trajectories is possible for such systems, but the quality of results is still limited by both data and computational methods. Specifically, it is feasible to learn high quality models from the best available algorithms from data sources already experimentally feasible for these systems, but improved data that pushes the boundaries of what is experimentally feasible does lead to better fits especially with respect to rare intermediate species. Furthermore, the need for better algorithms is very dependent on the data, with the quality of the final results highly algorithm dependent for poorer data sources but largely algorithm-independent for richer data sources.

Comparing the three optimization methods, we may conclude that SNOBFIT does better than the other two when it comes to fitting parameters of our virus capsid assembly model. MCS is known to be a superior method to SNOBFIT for some other applications (33) but SNOBFIT appears to be particularly well suited to dealing with the high stochastic noise typical of our data fitting problem. We note that this stochastic noise is inherent to the fact that we are using an SSA model to sample trajectories, a decision that itself has proven necessary to sampling large enough numbers of trajectories in reasonable amounts of time. This same issue would be expected to confront any simulation-based optimization of a system faced with similar combinatorial blowup of intermediate species, a general issue of self-assembly models but also one confronting other systems for which similar rule-based modeling have been applied (21-27). We would therefore argue that our observations are likely to be far more broadly applicable than just fitting capsid assembly models. On the other hand, SNOBFIT tends to identify a larger uncertainty in fits than do the other methods. We believe in these cases, the other methods are probably underestimating their true uncertainty since they do not survey the parameter space as thoroughly. The Kumar method largely relies on local optimization and thus might be expected to miss near-optima that also yield plausible fits to the true data. MCS is intended to be a global optimizer like SNOBFIT but might be expected to do a less complete survey of near optima in the presence of noise, explaining why it yields intermediate estimates of variance between the Kumar method and SNOBFIT. The difference among the inferred free energies from the application of the three methods is modest, considering the relatively small range of RMSDs between the methods (Table 2). However, some prior studies on synthetic datasets have shown that even a small variance in inferred parameters and free energies may lead to drastic changes in intermediate distributions and assembly pathways (7-9,11). It remains an open question whether curve fitting in true viral systems with older SLS data is sufficiently precise to derive accurate pathways and intermediate distributions in the face of imprecision in inferred rate constants or whether multiple substantially different pathway sets might be consistent with the observed data (9).

One must be cautious in considering such conclusions definitive, however, because of the dependence of the results largely on synthetic data. The need for synthetic data, rather than solely real experimental measurements, is largely due to 3 concerns: 1) we cannot rigorously evaluate accuracy of fitting without a known ground truth, which is unavailable for the real system; 2) we are interested in understanding the limits theoretically possible for these approaches and thus explore an idealized model rather than any true data as our representation of maximally data-rich experimental data; and 3) with synthetic data, we can work in a parameter domain in which trajectories are qualitatively similar to those of the real system but much faster to simulate, allowing us to test the full potential of the algorithms rather than relying on heuristic compromises needed with the real HBV data. Point 1) is the most difficult issue to sidestep but is also the major reason work in this direction is important: at present, there is to our knowledge no alternative approach to our data-fitting methods for learning detailed kinetic models of a non-trivial self-assembly system.  Point 2) merits further study using a variety of real data sources, including NCMS (37), DLS (38), and SAXS (39) that should provide information somewhere between the 3-SLS used in our current practice and our idealized model of NCMS.  While all of these methods have been used for capsid assembly studies previously, we are not aware of any one system for multiple such data types are available to allow a fair comparison and believe it is wisest to identify likely best practices by purely *in silico* studies like that published here before committing to a major experimental undertaking.  Point 3) is an issue of compute power and is in principle solvable by applying more powerful computers for longer times than we have available.  To drop the heuristic compromises on the real data for our current cluster hardware (typically consisting of 80 compute nodes in continuous use), however, would require years of continuous compute time and is therefore achievable in principle but not in practice.  The kinds of resources that direction would require exist but we again believe it would be wiser to identify best practices with these compromises to either find ways to bring down the cost or develop a clearer justification for a major commitment of compute hardware to this task.

While the results presented here must be taken with some caution, they nonetheless suggest that there is tremendous potential to simulation-based data fitting as a way of solving problems in macromolecular assembly that are not amenable to any purely experimental or purely theoretical technologies currently available.  Our results show that such methods can in principle extract far more information from feasible experimental data sources than is generally recognized.  While there is much to be done in learning the limits of these methods and establishing best practices for their use, there is strong reason to believe they can be a transformative technology for understanding macromolecular assembly processes and for the much broader project of developing predictive quantitative models of complex systems in biology.

# Author Contributions

L.X. and R.S. conceived the study and designed the experiments.  L.X. wrote the novel code developed for the study.  L.X. performed the simulation experiments.  L.X. and G.R.S. contributed analytical tools.  L.X., G.R.S., and R.S. analyzed the data. L.X., G.R.S., and R.S. wrote the manuscript.

# Acknowledgments

We thank Dr. Nikolaos Sahinidis for his advice on derivative free optimization methods. This work was supported by U.S. National Institutes of Health awards No. 1R01AI076318 (to L.X., G.S., and R.S.) and No. 1R01CA140214 (to R.S.), the Computational Biology Department, Carnegie Mellon University (to L. X.) and Department of Biological Sciences, Carnegie Mellon University (to G.S.).

# Appendix A: Supplementary Figures

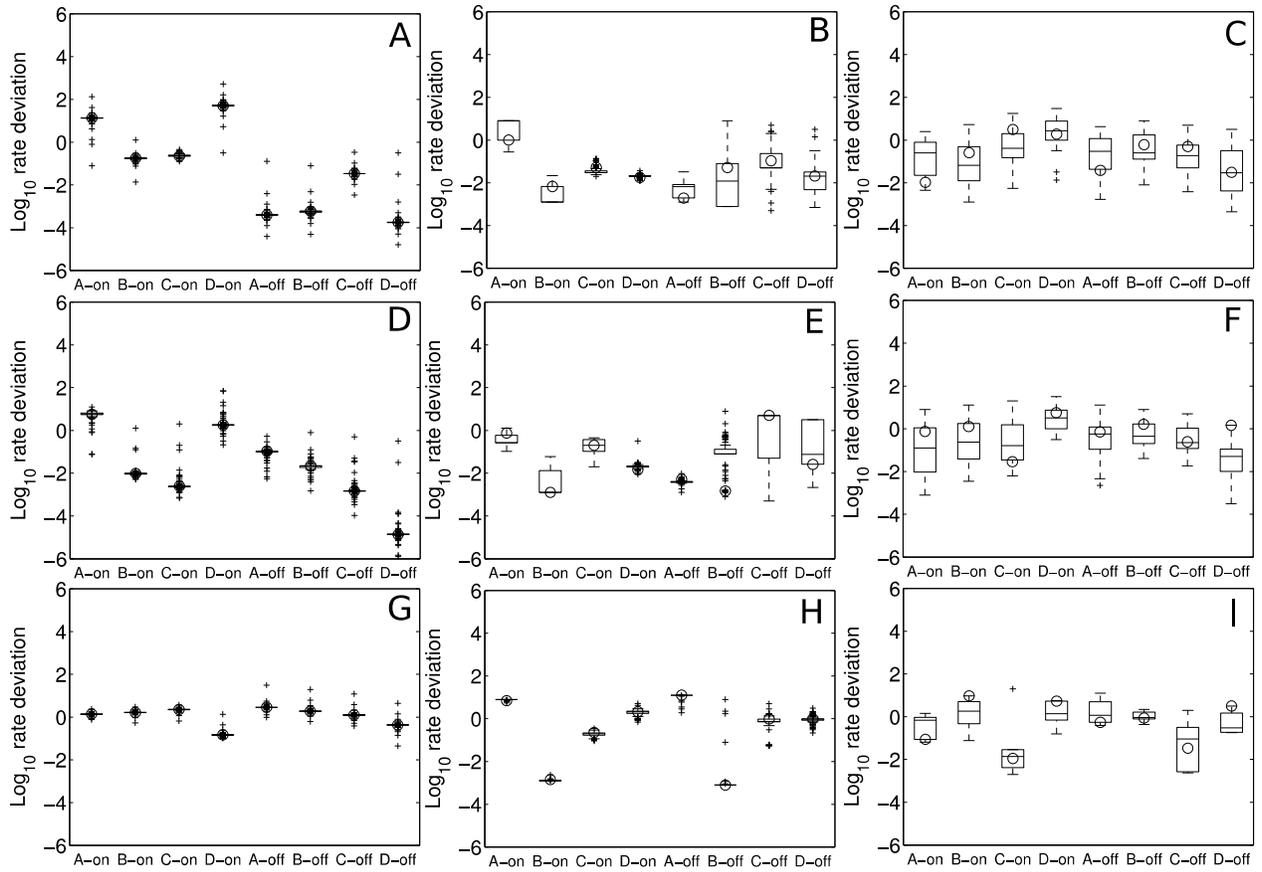

**Supplementary Figure S1. Deviation of inferred log parameters from true values on the synthetic data sets for each algorithm and data type.** In each plot, circles mark the deviation with minimum RMSD, and boxes mark mean and range of variation for high-quality fits, defined to be those whose RMSD is within two standard deviations of the estimated noise level. The eight parameters in log space on each plot correspond to A on-rate, B on-rate, C on-rate, and D on-rate in $M^{-1}s^{-1}$ followed by A off-rate, B off-rate, C off-rate, and D off-rate in $s^{-1}$. The subfigures show the three algorithms separated by column (A, D, G) Kumar; (B, E, H) MCS; (C, F, I) SNOBFIT and the three data sources separated by row (A, B, C) 1-SLS; (D, E, F) 3-SLS; (G, H, I) NCMS.

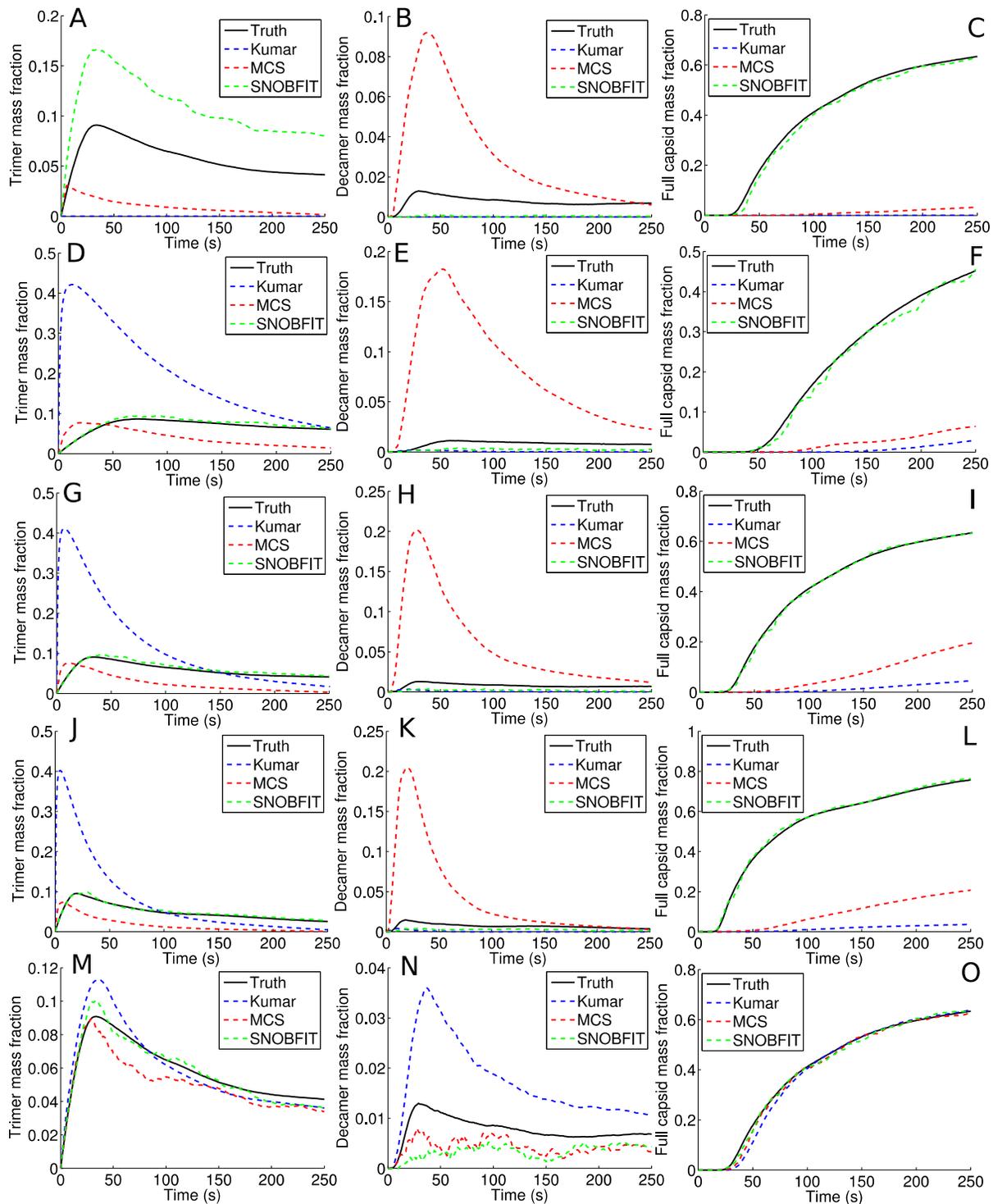

**Supplementary Figure S2. Best fit curves from synthetic data for a representative sample of mass fractions.** Each subfigure compares true mass fraction in solid black to the best fits for the Kumar method (dashed blue), MCS (dashed red), and SNOBFIT (dashed green). Columns represent mass fractions for a high-abundance trimer species (A, D, G, J, M), a low-abundance decamer species (B, E, H, K, N), and complete capsid (C, F, I, L, O). Rows represent data sources: (A, B, C) 1-SLS; (D, E, F) 5.3 µM, (G, H, I) 8.0 µM, and (J, K, L) 10.6 µM curves from 3-SLS; and (M, N, O) NCMS.

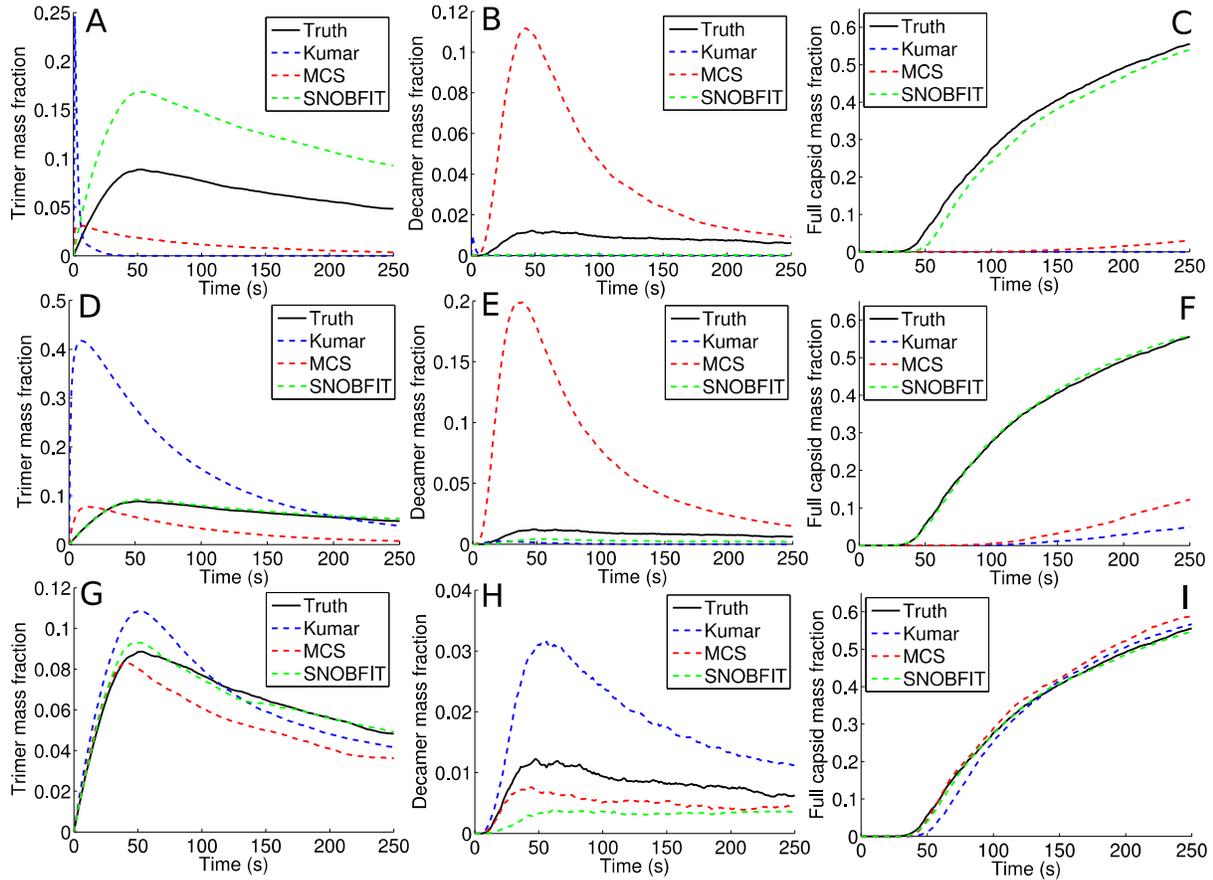

**Supplementary Figure S3. Predicted versus true mass fractions for synthetic data by each method and data source for a concentration not used in data-fitting for a representative selection of intermediate species.** Each subfigure compares true SLS in solid black lines versus the predictions of the Kumar method in dashed blue, MCS in dashed red, and SNOBFIT in dashed green for one data source at a concentration omitted from data-fitting. Rows correspond to data source: (A, B, C) 1-SLS; (D, E, F) 3-SLS; (G, H, I) NCMS. Columns to intermediate species profiled: (A, D, G) medium-abundance trimer intermediate; (B, E, H) low-abundance decamer intermediate; (C, F, I) high-abundance complete capsomer.